\documentclass[letterpaper,twocolumn,10pt]{article}
\pdfoutput=1
\usepackage[margin=1in,headsep=0pt]{geometry}
\usepackage{graphicx}
\usepackage{url}
\usepackage[labelfont=bf]{caption}
\usepackage{parskip}
\usepackage{microtype}
\begin{document}

\date{}

\title{\Large \bf Implementing BOLA-BASIC on Puffer: \\Lessons for the use of SSIM in ABR logic}

\author{
\rm
\setlength{\tabcolsep}{12pt}
\def\arraystretch{1.05}
\begin{tabular}{cccc}
  Emily Marx & Francis Y. Yan\textsuperscript{$\dagger$} & Keith Winstein \\
\end{tabular}\\[3.5ex]
  {\it Stanford University, \textsuperscript{$\dagger$}Microsoft Research}
}

\maketitle

\begin{abstract}

One ABR algorithm implemented on Puffer\footnote{\url{https://puffer.stanford.edu}} is BOLA-BASIC, the simplest
variant of BOLA. BOLA finds wide use in industry, notably in the
MPEG-DASH reference player used as the basis for video players at 
Akamai, BBC, Orange, and CBS~\cite{bola}.
The overall goal of BOLA is to maximize
each encoded chunk's video quality while minimizing rebuffering. To
measure video quality,  Puffer uses the structural similarity metric
SSIM, whereas BOLA and other ABR algorithms like BBA, MPC, and Pensieve
are more commonly implemented using bitrate (or a variant of bitrate).

While bitrate is frequently used, BOLA allows the video provider to
define its own proxy of video quality as the algorithm's ``utility''
function. However, using SSIM as utility proved surprisingly complex for
BOLA-BASIC, despite the algorithm's simplicity. Given the rising
popularity of SSIM and related quality metrics, we anticipate that a
growing number of Puffer-like systems will face similar challenges. We
hope developers of such systems find our experiences informative as they
implement algorithms designed with bitrate-based utility in mind.

\end{abstract}

\maketitle

\section{Background: SSIM and video quality}
ABR algorithms like BOLA decide the bitrate at which to download each chunk
of video into the client's buffer. As the throughput of the network varies, the
goal is to minimize stalls caused by an empty buffer while choosing
high-quality encodings (see the 
Puffer paper~\cite{puffer} for more background on
ABR). Much recent work has focused on objective metrics for an
encoding's subjective quality. Measuring perceived video quality is
difficult, and no metric is perfect. Recently, Netflix's VMAF has
attempted to combine the strengths of several metrics by using them as
input to a machine-learning model~\cite{vmaf}.

In the BOLA paper and DASH player, the ``utility'' metric that
measures video quality is the normalized logarithm of encoded chunk
size. Using size as utility is effectively equivalent to using bitrate
and is a common choice. Bitrate is also easily computed, and is already
made available to the ABR logic in DASH and other streaming
protocols~\cite{bitrate}.
However, video quality does not correlate directly with
bitrate~\cite{bitrate}. When
the content of a video changes over time, like the live TV streamed on
Puffer, some chunks are easier to encode than others. For instance, a
stream may have a black screen for several seconds, followed by a
fast-moving scene. The all-black chunks can be encoded more efficiently
than the scene, so they are likely to both look better \emph{and} be smaller
than more complex chunks.

SSIM~\cite{ssim}
is one of many metrics that attempt to capture this variation.
Unlike bitrate, SSIM is a compute-intensive metric, and Puffer is
unusual in providing it at runtime. In fact, 18 of Puffer's cores are
solely devoted to SSIM. Many systems use metrics like SSIM, or the older
PSNR, to evaluate chunks \emph{after} they are chosen by the ABR algorithm.
However, to our knowledge Puffer is the first study to use SSIM or PSNR
as a factor in the ABR decision.

\section{BOLA-BASIC on Puffer}
Unlike many other ABR
algorithms, BOLA-BASIC is very simple in implementation. The objective
is a function of each encoded chunk's utility and size, along with the
buffer level and two control parameters. If the buffer is not too full,
the algorithm chooses the encoded chunk with the highest objective.
Otherwise, BOLA elects not to download any chunk before the next
decision opportunity~\cite{bola}.

The first implementation of BOLA-BASIC on Puffer was a direct
representation of the algorithm as presented in the
paper~\cite{bola}, using SSIM
in decibels (SSIMdB) as the utility function. Puffer also uses SSIMdB to
measure video quality in the other ABR algorithms implemented on the
platform, as well as to measure their performance (in conjunction with
stall ratio).

\section{A less basic BOLA-BASIC}
The BOLA authors shared
their expertise to help optimize the initial implementation of
BOLA-BASIC. This resulted in three changes.

The first two changes related to BOLA's decision thresholds, i.e.
the buffer levels at which BOLA changes its bitrate decision. At the
first threshold, BOLA switches from choosing the smallest to the
second-smallest encoding. At the last threshold, BOLA switches from the
largest chunk to no chunk at all.

\begin{figure}[h]
\centering
\includegraphics[width=1.0\columnwidth]{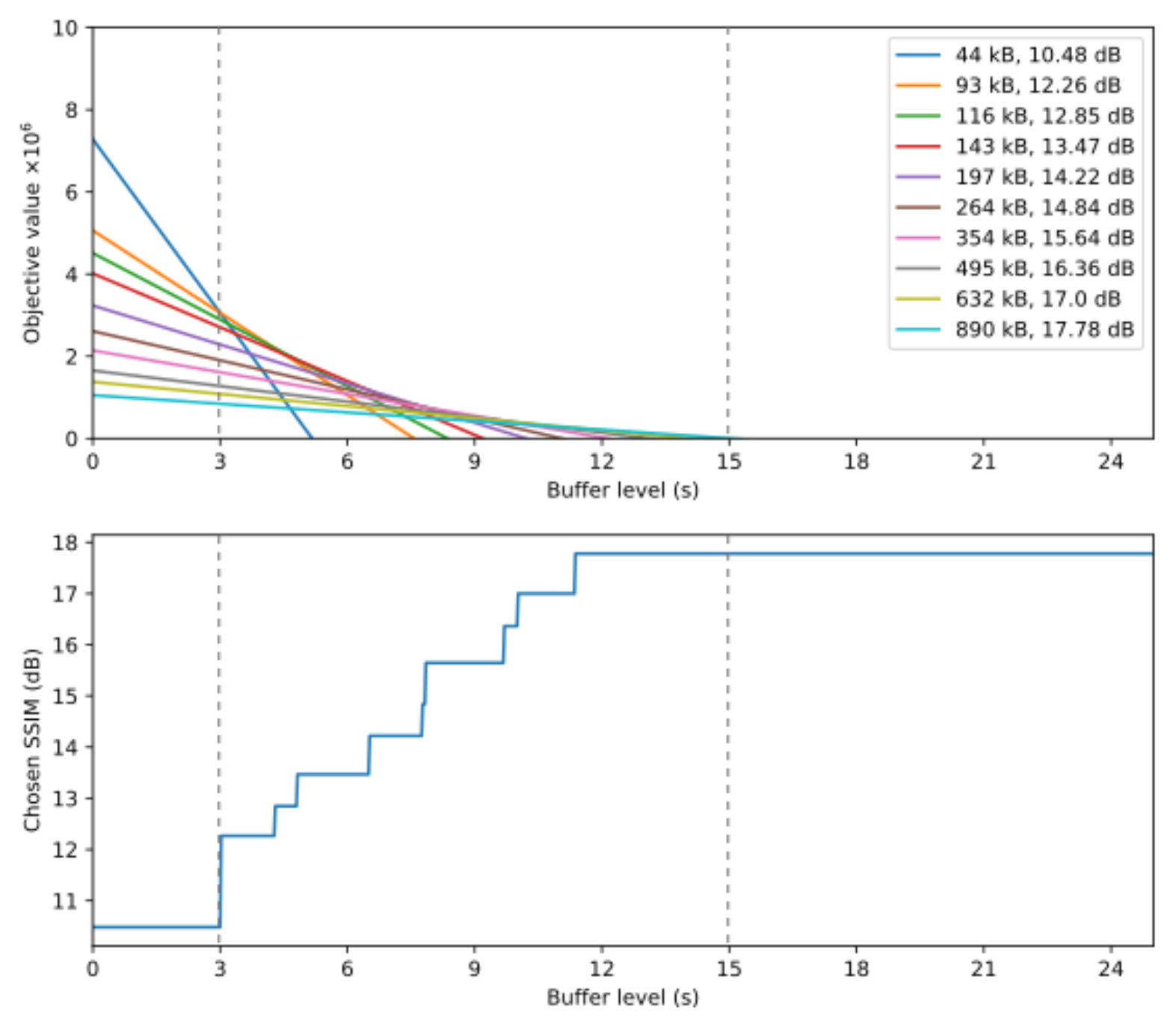}
\caption{Decision thresholds for average SSIM/sizes in
BOLA-BASIC ``v1''. Dashed lines mark min/max buffer level. Thresholds
beyond max buffer are hypothetical.}
\label{fig:average-objectives-decisions-v1}
\end{figure}

The decision thresholds in the average case are shown in Figure~\ref{fig:average-objectives-decisions-v1}.
BOLA's control parameters $V$ and $\gamma$ are calculated statically, using
long-term averages for the utility and size of each of the ten encoded
formats. So, Figure~\ref{fig:average-objectives-decisions-v1} shows the decisions BOLA would make if the ten
options produced by the encoder at some time slot exactly matched the
averages used to calculate the parameters. Puffer uses 3 and 15 seconds
for minimum and maximum buffer size, which correspond to the first and
last thresholds in the average case.

\begin{figure}[ht]
\centering
\includegraphics[width=1.0\columnwidth]{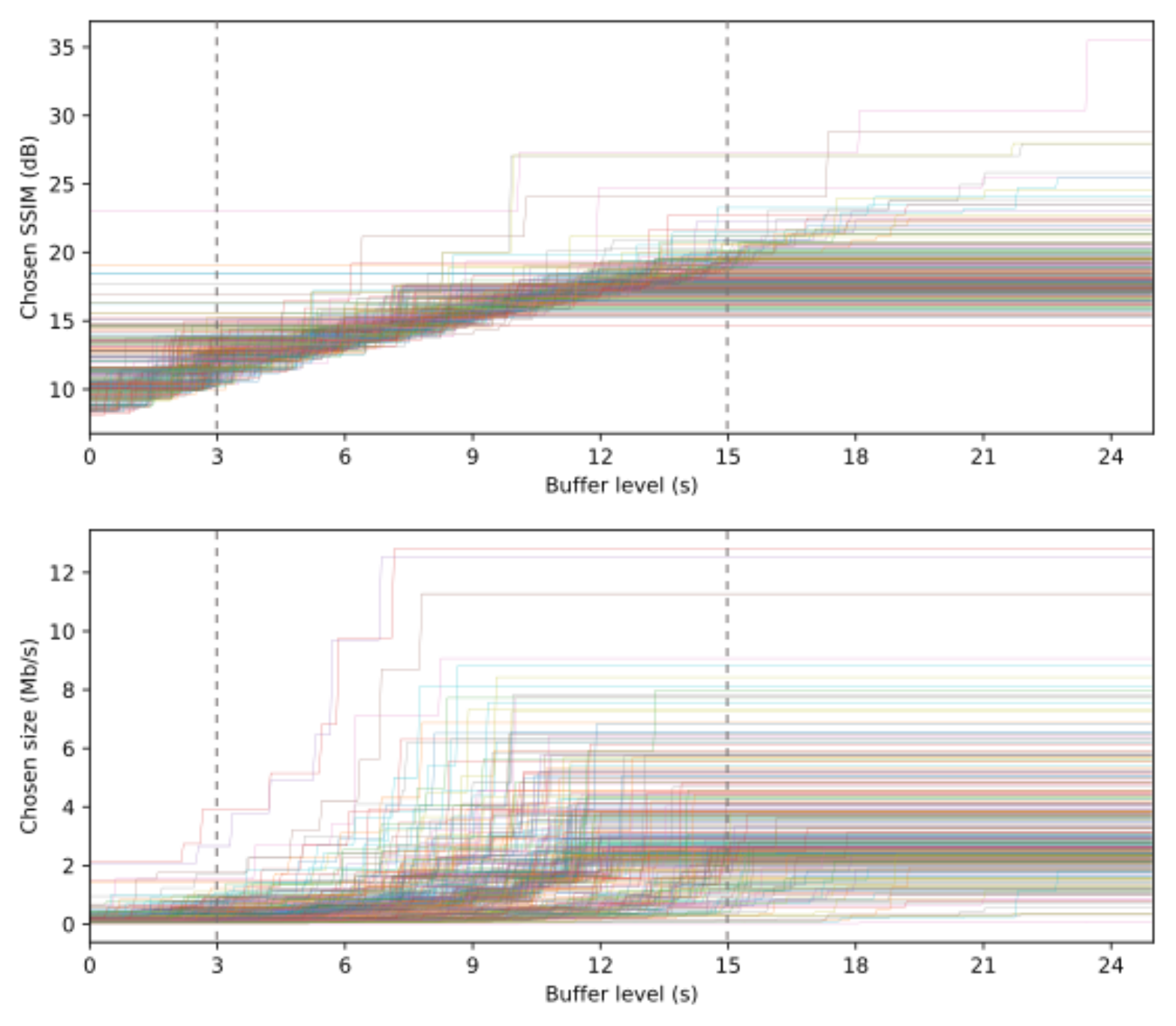}
\caption{Decision thresholds for BOLA-BASIC ``v1'' over a
10-minute clip. Each line represents a single 120-frame chunk. Dashed
lines mark min/max buffer level. Thresholds beyond max buffer are
hypothetical.}
\label{fig:cbs-decisions-v1}
\end{figure}

Figure~\ref{fig:cbs-decisions-v1} shows the thresholds over a 10-minute CBS clip, using the
naive implementation of BOLA-BASIC. Each line corresponds to the
encodings produced for one chunk. Notice that many of the thresholds
appear above the maximum buffer level of 15 seconds.

Since BOLA elects not to send a chunk at buffer levels above the
maximum, BOLA will never exercise any of the thresholds beyond 15
seconds. To mitigate the appearance of such thresholds, the authors
suggested two modifications to the utility function.

\begin{figure}
\centering
\includegraphics[width=0.75\columnwidth]{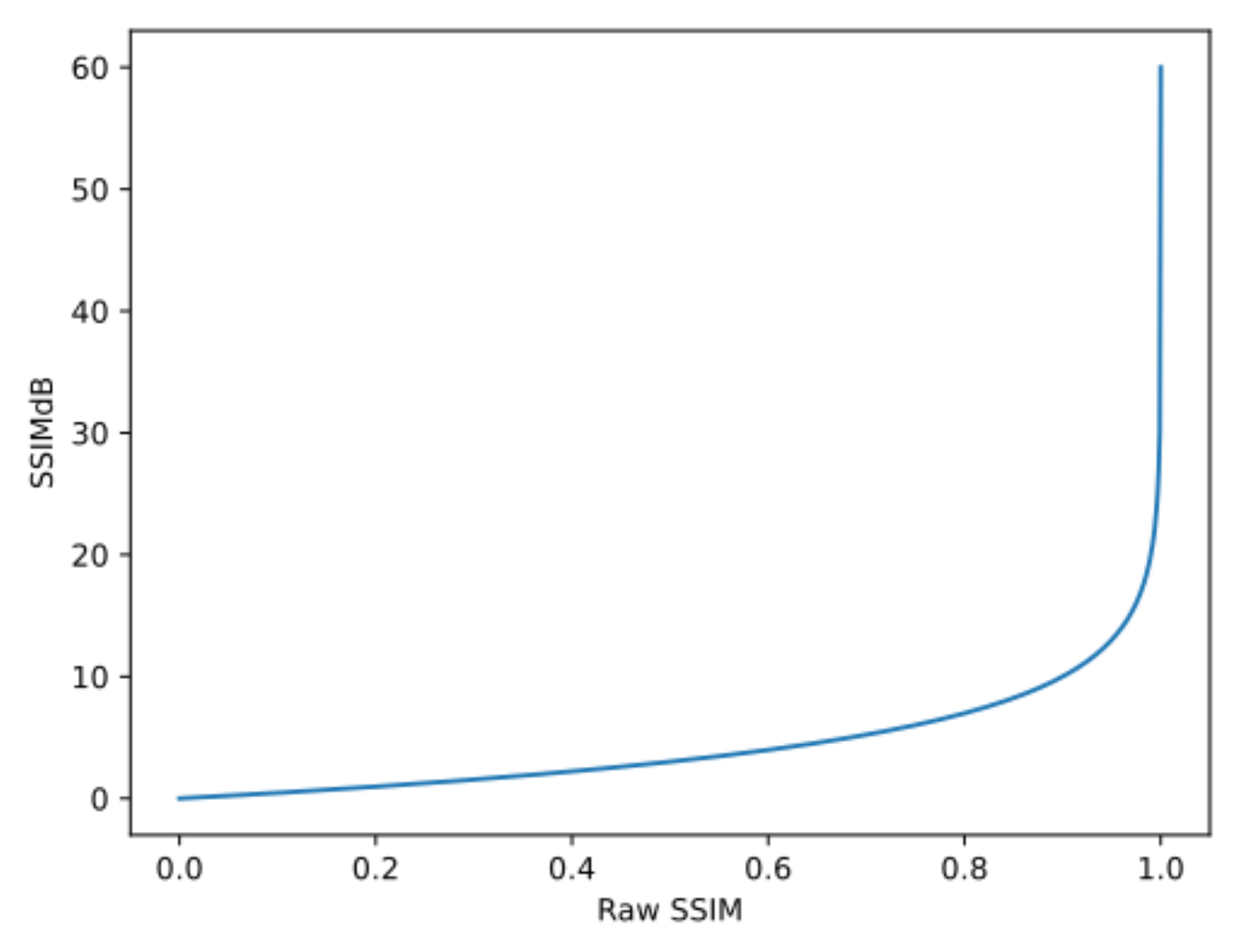}
\caption{Conversion of SSIM to decibels is asymptotic in
the upper range.}
\label{fig:SSIM-dB-vs-raw}
\end{figure}

The first change in BOLA-BASIC ``v2'' is to use the raw value of SSIM,
without converting to decibels. Although the data correlating SSIMdB
with perceived quality is noisy, there is some evidence that the decibel
transformation creates an approximately linear relationship with human
preference~\cite{salsify}.
However, the transformation to decibels expands the values
in the upper SSIM range, as shown in Figure~\ref{fig:SSIM-dB-vs-raw}.

As a result, the utility of higher-quality encodings is inflated
using SSIMdB relative to SSIM, pushing the thresholds to the right. So,
the BOLA authors expected the algorithm to behave better with raw SSIM.

As a second adjustment to the utility function, the authors
suggested using the maximum \emph{possible} utility (i.e. 1.0 for raw
SSIM) in the parameter calculations, rather than the maximum
\emph{average} utility. On average, the highest-quality encoding
available is 0.983, so using 1.0 instead is a significant change.

\begin{figure}[h]
\centering
\includegraphics[width=1.0\columnwidth]{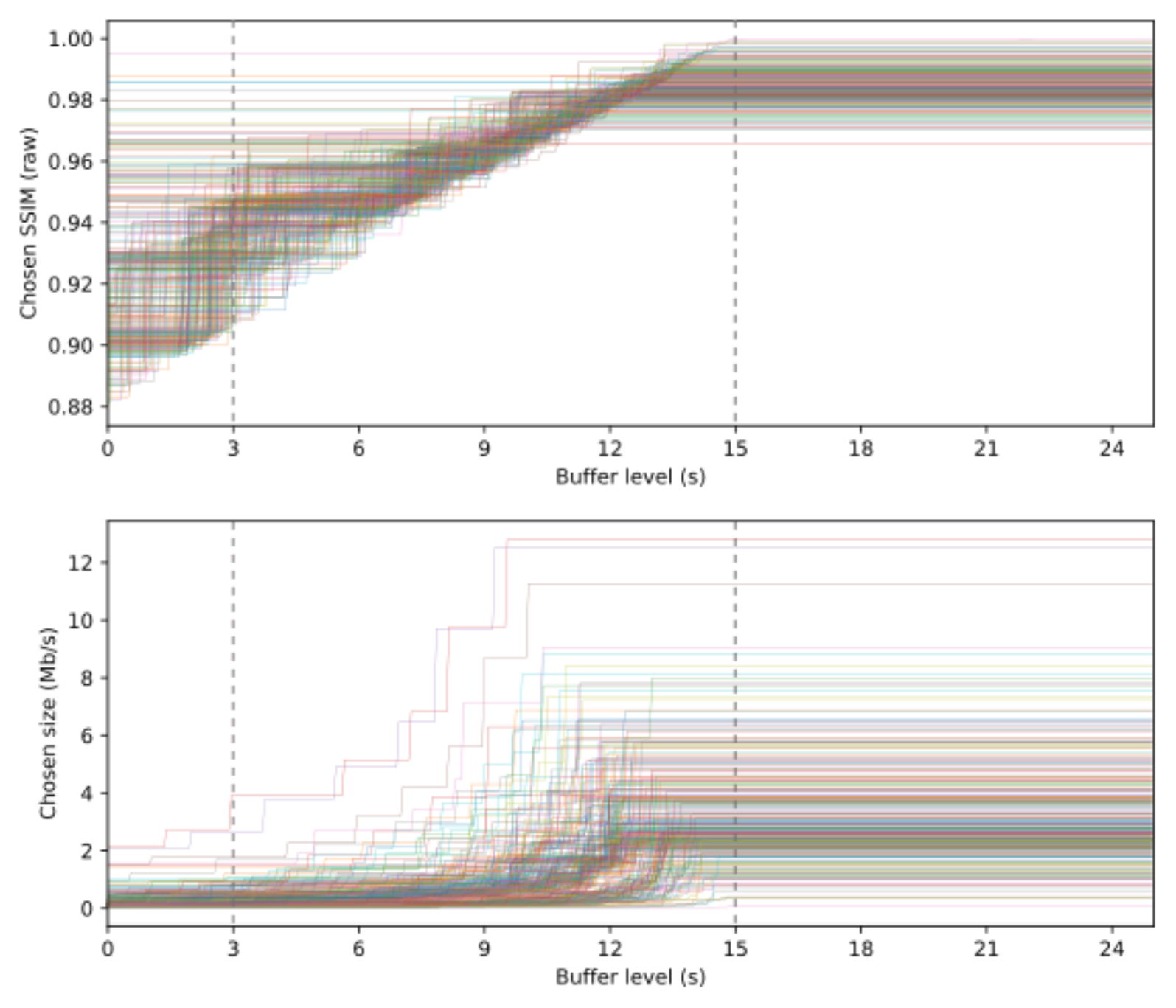}
\caption{Updated version of Figure~\ref{fig:cbs-decisions-v1} for
BOLA-BASIC ``v2''.}
\label{fig:cbs-decisions-v2}
\end{figure}

As shown in Figure~\ref{fig:cbs-decisions-v2}, most of the decision thresholds shift below
the maximum buffer size after making these two changes.

Separately, the authors suggested a third change, relating to the
case where the objective is negative for all available encodings. In
this situation, the DASH reference implementation pauses until the
buffer has drained enough so that some objective is non-negative. Puffer
implements ABR on the server, so this pause was simulated to avoid
introducing BOLA-specific logic in the server. Specifically, the authors
suggested that if all objectives are negative, BOLA-BASIC ``v2'' should
choose the chunk with highest utility rather than highest objective.
This simulates the client's pause, since the point at which the
objective becomes positive is a factor only of utility, not the size in
the denominator (see Figure~\ref{fig:objective-function}).

\section{BOLA-BASIC and SSIM}
The changes in the second
implementation of BOLA-BASIC on Puffer do not address a fundamental
issue with the use of SSIM (whether decibels or raw) in BOLA. The BOLA
parameters are calculated statically, but SSIM varies dynamically with
bitrate. According to the paper, BOLA can use any utility function as
long as the utility of the available encodings for each chunk is
nondecreasing with respect to their size. In fact, the paper identifies
the ability to define utility in ``very general'' ways as a unique
strength of BOLA. SSIM satisfies BOLA's requirement of monotonicity
with respect to size \emph{within} the encodings for each chunk (120 frames on
Puffer). However, as discussed above, utility can vary independently of
size, and therefore bitrate, \emph{across} chunks.  

Unlike direct functions of bitrate, SSIM captures this variation.
However, this causes BOLA to behave very differently when choosing
between a set of format options with low utility than when choosing
between higher-quality chunks. Due to the way utility is used in BOLA's
objective (Figure~\ref{fig:objective-function}), when utility is near $-\gamma p$, a small gain in utility
can outweigh a large bitrate increase, particularly if the buffer is
near empty. For instance, BOLA could prefer a format twice as large
offering only a 0.4dB SSIM increase. While downloading this much larger
chunk, the near-empty buffer may drain completely. In contrast, when
choosing between chunks whose utility is larger in magnitude relative to
$-\gamma p$, the same utility gain has less impact relative to the bitrate
difference.

\begin{figure}[h]
\centering
$$\frac{V(v_m + \gamma p) - Q(t_k)}{S_m}$$
\caption{BOLA's objective. $V$ and $\gamma$ are control
parameters, $p$ is chunk duration, $Q(t_k)$ is buffer level,
$S_m$ is size, and $v_m$ is utility.}
\label{fig:objective-function}
\end{figure}

It seems that the static parameter
calculation inherent to BOLA is fundamentally incompatible with a
utility function involving variables beyond bitrate. It's
interesting that the variation of SSIM with bitrate is a strength
of SSIM as a utility metric, but makes SSIM less amenable to
algorithms based on more approximate utility metrics.

\section{Results}
\begin{figure}[h]
\centering
\includegraphics[width=1.0\columnwidth]{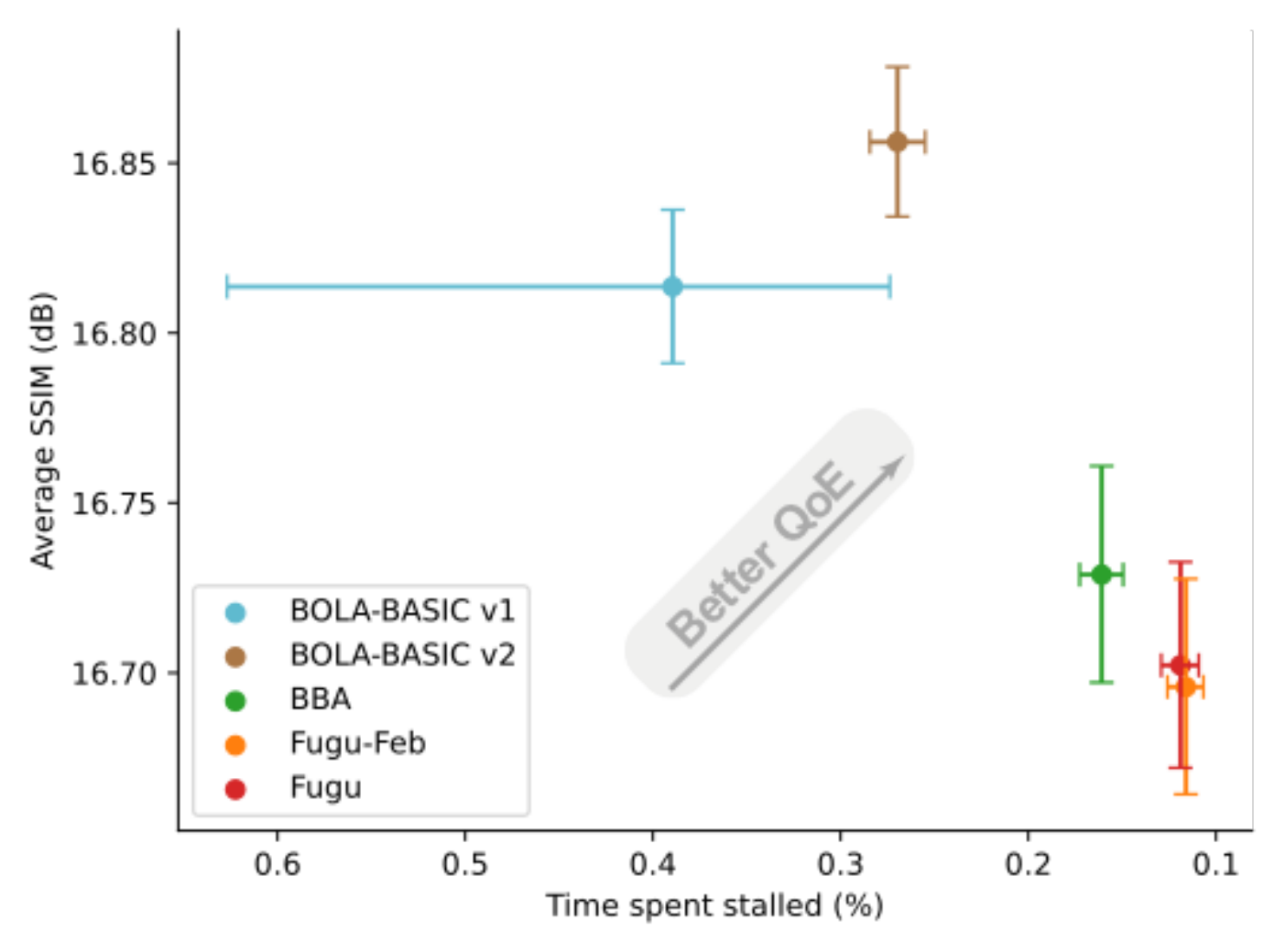}
\includegraphics[width=1.0\columnwidth]{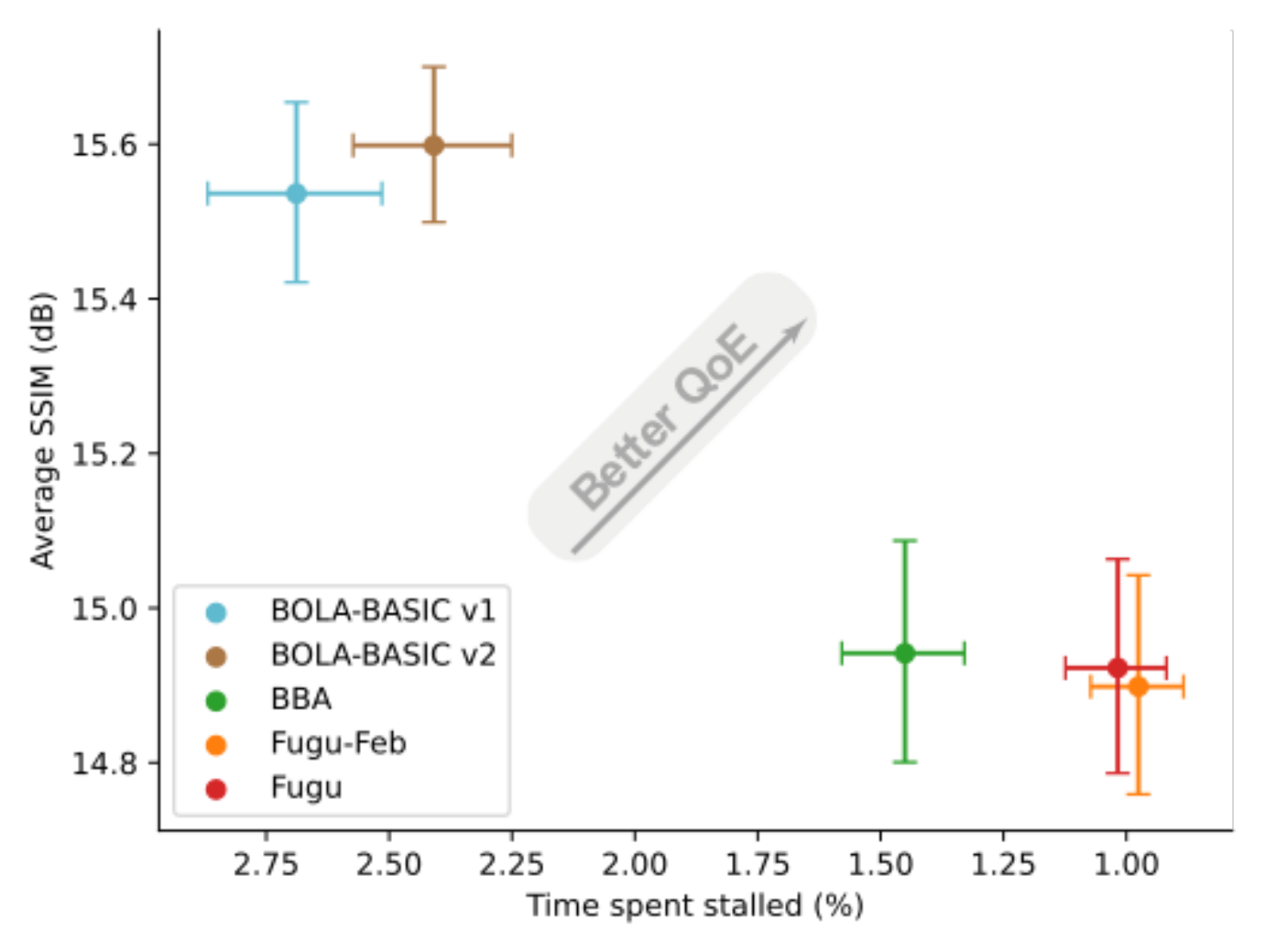}
\caption[Caption for LOF]{Performance of both versions of BOLA-BASIC on Puffer (95\%
confidence intervals), for all stream speeds (top plot) and slow
streams only (bottom plot). Both plots show data from 2020-07-26 to
2020-11-07 (see Puffer\protect\footnotemark ~for latest
data). All speeds comprises 685,022 streams (17.7 stream-years);
slow speeds comprises 96,646 streams (1.5 stream-years).}
\label{fig:results}
\end{figure}

Figure~\ref{fig:results} shows the performance of the
initial ``v1'' implementation of BOLA-BASIC, as well as the ``v2''
implementation with the three changes discussed above. Also shown are
BBA and two ML-based ABR algorithms developed on Puffer. Each of the
five algorithms has over 3.5 cumulative stream-years of data.

\footnotetext{\url{https://puffer.stanford.edu/results}}
The two versions of BOLA-BASIC are broadly similar at this
timescale, with ``v2'' showing slight improvement in SSIMdB and stall
ratio. Relative to BBA, both versions improve SSIMdB while increasing
stall time.

\section{Conclusions}
It should be noted that BOLA-BASIC is considerably simpler than
production implementations of BOLA, e.g. BOLA-E and BOLA-O in the DASH
reference player. Also, Puffer has several architectural differences
from DASH and other client-side players (see the
Puffer paper~\cite{puffer} for
detail). For these reasons, we don't intend to present these results as
a comment on the general performance of BOLA relative to the other
algorithms. Instead, we see these results as an indication of the
surprising complexity possible when using SSIM in even the simplest ABR
algorithms.

Newer video quality metrics like SSIM more accurately reflect human
perception. However, existing ABR algorithms may not always be designed
to take full advantage of these metrics. If measurements like SSIM are
the future of ABR, Puffer will not be alone in facing these challenges.

\section{Acknowledgements}
Many thanks to the BOLA authors for their extensive advice. 

\label{beforerefs}

\bibliographystyle{ieeetr}
\bibliography{reference}

\end{document}